# On SH Molecules in Umbral Spectra

K. Sinha

*(ksinha2000@hotmail.com)*

*Aryabhatta Research Institute of observational sciencES, Manora Peak, Nainital - 263129, India.*

**Abstract :** Following our predictions on detectability, very weak lines of the SH molecules have been reported as identified in the photospheric spectrum of the Sun. This could be the first detection of a mercapto radical in the solar spectrum, thus placing confidence in the solar abundance of Sulfur. The observations additionally tested the utilized theoretical band oscillator strength.

Sunspots being cooler than the photosphere are the hosts to a large number of molecular species. However, owing to photospheric radiation scattered into the observed umbra, the discoverers could not detect the lines of SH in the sunspot spectrum where they are expected to show larger than the photospheric equivalent widths (EWs in short). Detection of weak lines in photospheric spectrum coupled with a no-detection of the relatively strong same lines in the sunspot spectrum might cause doubts on the detection itself.

The above problem is investigated here in detail with a choice on photospheric and sunspot models and micro-turbulence values. The new results indicate that the predicted sunspot lines are about half intense than reported before and the lines remain of detectable strengths. Further, a need for a laboratory determination of oscillator strengths is felt.

## 1. Introduction

The Sulfur abundances are known to be sensitive to the metallicities in stars and therefore useful to evaluate the chemical evolution of the Galaxy.

The two atomic lines at 869.4 nm and at 1082 nm have been utilized to determine solar abundances of Sulfur with the help of 1 D and 3 D solar models. For a brief review on the solar abundances the paper by Asplund et al (2009) may be consulted.

Since a molecule presents several lines in a considered region of the spectrum, the same may also be used to determine abundances in a stellar atmospheres. For some time, the molecule SH was expected to be present in the solar spectrum (Sinha, 1990). In a detailed investigation on the Sulfur containing solar photospheric molecules SH, CS, SO, NS and $S_2$, Sinha et al (1976, 77) concluded that the abundances of these molecules decreased successively by an order of magnitude and also that very weak lines of the molecule SH could be identified in the photospheric spectrum. Additionally, the same molecule was said detectable in the IR spectra of sunspots. Joshi and Pande (1979) carried out the investigations further in Zwaan's (1974) sunspot model and finding large EWs these authors suggested that the ultraviolet lines belonging to the A – X transition of the molecule are detectable and that they could also be utilized to determine the isotopic abundances of Sulfur.

With the help of a high resolution UV spectra and synthetic spectrum for the solar photosphere, Berdyugina and Livingston (2002) identified the molecule for the first time in the solar photosphere and derived an oscillator strength $f_{0,0} = 2.2 \times 10^{-3}$. However, as they state, their initial investigations for the strong umbral lines was unsatisfactory in view of stray photospheric light. In view of the importance of the problem, we decided to revisit the problem with the help of a range in photospheric and sunspot models and fresh data on partition functions and dissociation constants.

## 2. Formulations and Calculations

The computer code to calculate the EWs was modified for inclusion of partial pressures of a large number of atoms, the partition functions due to Irwin (1981) and the dissociation constants due to Sauval and Tatum (1984) and Tsuji (1973). The dissociation energy of SH was taken as 3.55 eV, abundance of Sulfur as N(S) = 7.21 on the usual log N(H) = 12.00 scale with the $^{32}S/^{34}S$ ratio as 22.565 and $f_{0,0}$ = 1.63 x $10^{-3}$ (Henneker and Popkie 1971).

In order to keep our investigations close to the previous ones, the 1 D photospheric and umbral model atmospheres were preferred. However, a choice of a more realistic 3 D model yields a lower Sulfur abundance 7.12 $\pm$ 0.03 (Asplund et al, 1999). It may be noted here that a choice of a different model requires a different elemental abundance and/or oscillator strength to match an observed equivalent width.

The chosen photospheric models for the calculations are those by Gingerich et al (1971), Holweger and Müller (1974), Vernazza et al (1976), Maltby et al (1986) and Grevesse and Sauval (1999). The chosen umbral models are those due to Henoux (1969), Zwaan (1974), Boyer (1980), Avrett et al (1981) and Stellmacher and Wiehr (1981). The calculations were carried out for all the $Q_2$ lines chosen by Sinha et al (1976, 77), all the five lines investigated by Berdyugina and Livingston (2002) and the six lines arising from J = 10.5 of the $P_1$, $P_2$, $Q_1$, $Q_2$, $R_1$ and the $R_2$ branches involved in the transition. The chosen micro-turbulence values are 0.85, 1.0 and 2.1 km $s^{-1}$.

### 3. Results and Discussions

The results of calculations for the solar photosphere are presented in Fig.1 which is in agreement with Berdyugina and Livingston (2004). The table indicates the minimum predicted equivalent width as 0.99 mÅ for the photospheric SH lines in the UV.

Fig. 2 thru 4 contain results for the sunspot models. Considering the good range in the sunspot model atmospheres, it is found that although compared to Joshi and Pande (1979), the predicted EWs are found lower by a factor 2, they are of detectable strengths.

The results of calculations in a tabular form can be obtained by sending an email to the author.

Thus, we are now faced with a situation where the stronger than photospheric SH lines are yet to be detected in the umbral spectrum. It is felt that in view of the unique requirement on isotopic determinations, efforts to detect the umbral lines are desired, may be in a clean region of the solar spectrum.

Also, there appears sufficient reason to reiterate that a laboratory determination of the oscillator strength of the ultraviolet transition in SH is required. In contrast to the theoretical $f_{0,0} = 1.63 \times 10^{-3}$ used here and reported by Henneker and Popkie (1971), Berdyugina et al (2002) derived $f_{0,0} = 2.2 \times 10^{-3}$ and Resende and Ornellas (2001) calculated the Einstein transition probability for the (0 – 0) band which yields $f_{0,0} = 2.896 \times 10^{-3}$. Any inaccuracies in the theoretical oscillator strengths are expected to be reflected in the abundance determinations and in the predicted EWs.

**4. Acknowledgement:** It is a pleasure to record here the valuable help received from my former student and friend Dr. C. S. Stalin who is a faculty now at the Indian Institute of Astrophysics, Bengaluru, India.

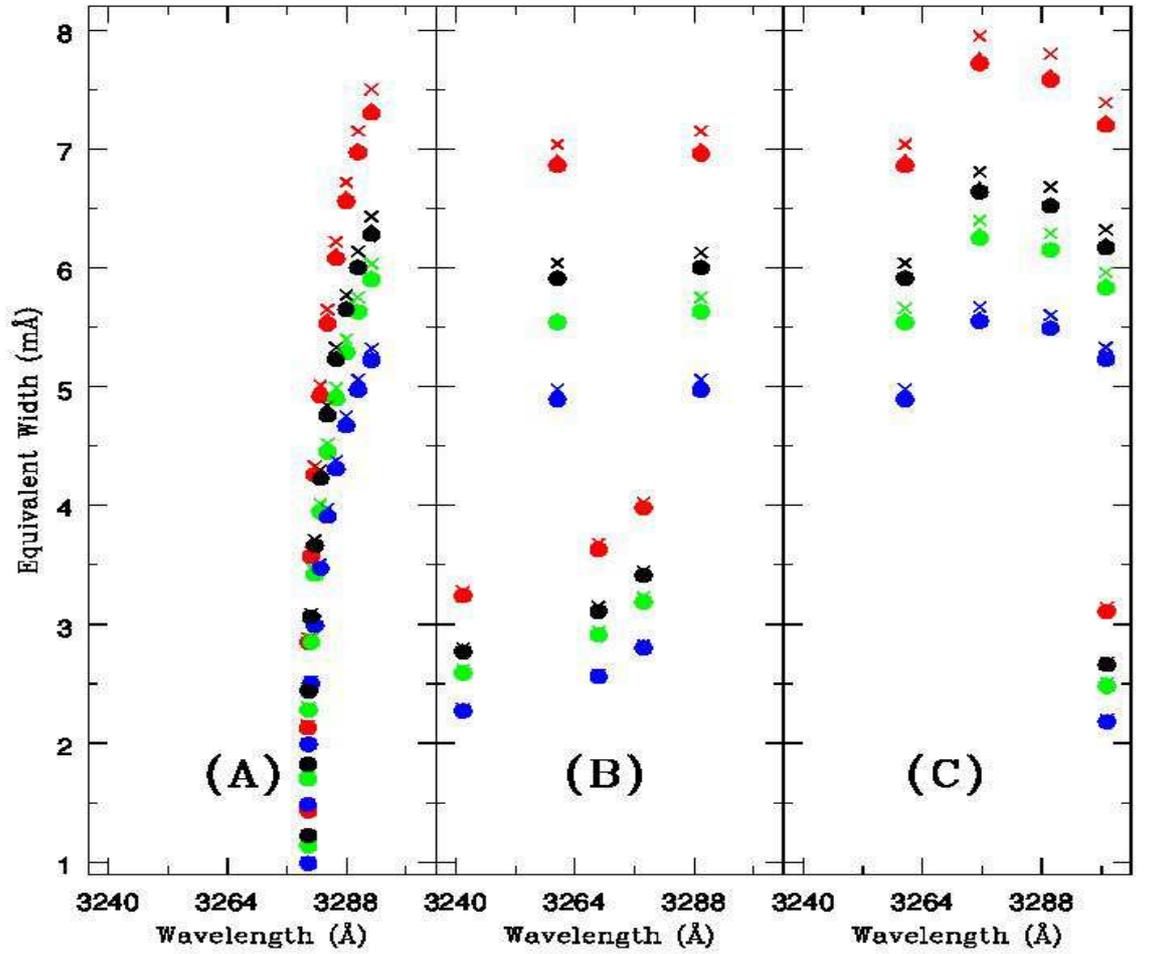

**Fig.1.** Predicted equivalent widths plotted against wavelengths in photospheric models with variations on micro-turbulence. Colour code : Red, Green, Blue and Black refer respectively to Gingerich et al (1971), to Grevesse & Sauval (1999), to Holweger & Müller (1974) and to Vernazza et al.(1976) photospheric models. Filled circles, filled triangles and crosses refer to micro-turbulence values as 0.85, 1.00 and 2.1 km s$^{-1}$ respectively. Panels (A), (B) and (C) refer respectively to (i) $Q_2$ lines arising from J = 1.5 to 11.5, (ii)one line each from the six branches with J = 11.5 and (iii)for the lines reported as present by Berdyugina & Livingston (2002).

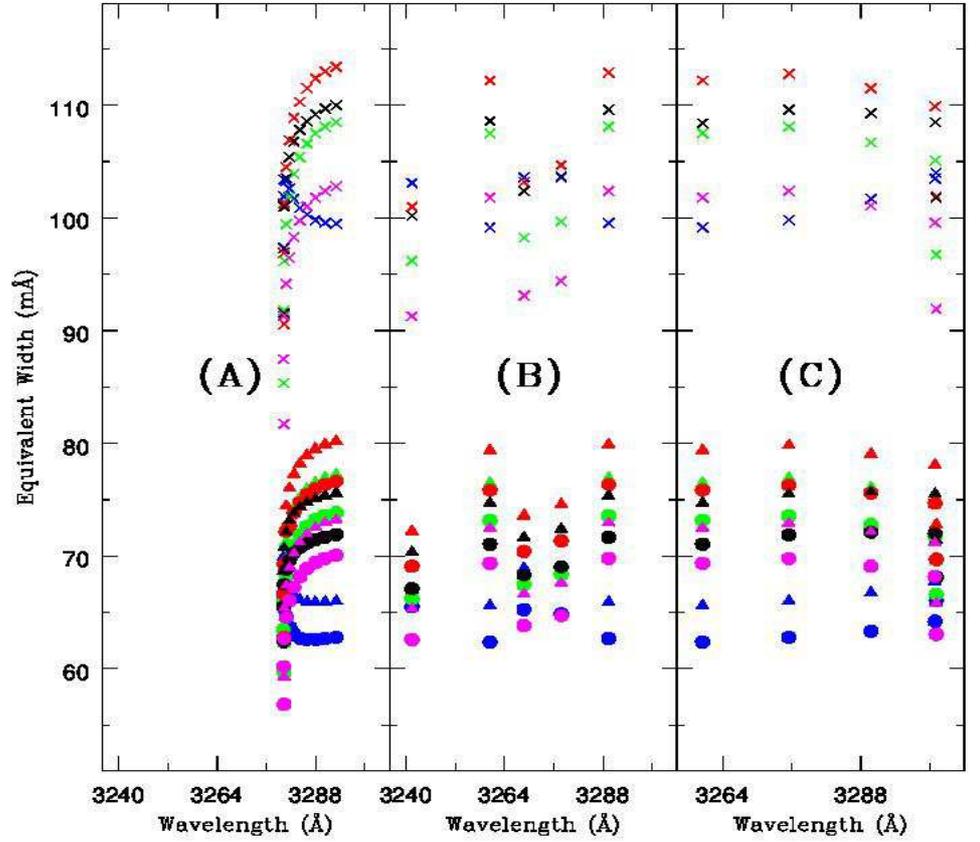

**Fig.2.** Predicted equivalent widths plotted against wavelengths in sunspot models with variation on micro-turbulence. Colour code : Red, Green, Black, Blue and Magenta refer to Stellmacher & Wiehr (1975), to Boyer 1980), to Avrett (1981), to Zwan (1974) and to Henoux 1969) sunspot models. Filled circles, filled triangles and crosses refer to micro-turbulence values as 0.85, 1.00 and 2.1 km s$^{-1}$ respectively. Panels (A), (B) and (C) are as described in Fig.1.

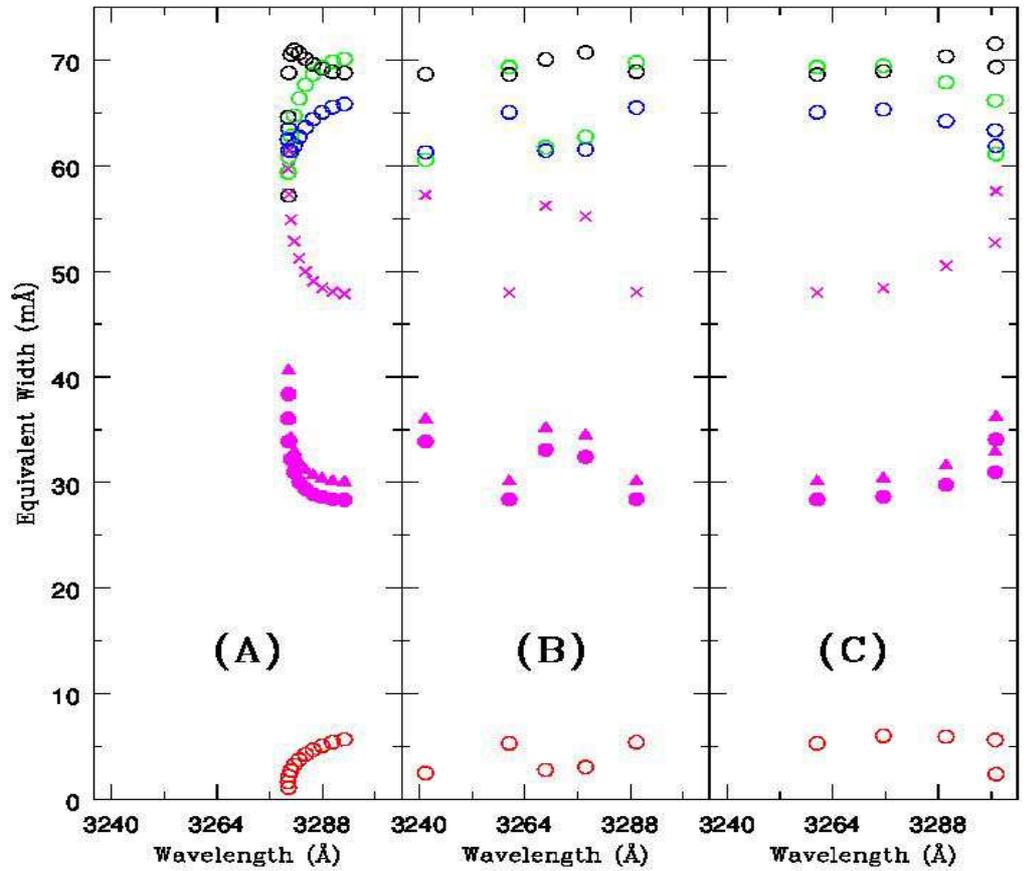

**Fig.3.** Predicted equivalent widths plotted against wavelengths in sunspot models due to Maltby et al. (1986). Color code : Red, Black, Green and Blue open circles refer respectively to the photospheric model and sunspot models at late, early and middle phases of solar activity. Filled circles, filled triangles and crosses in magenta refer to micro-turbulence values as 0.85, 1.00 and 2.1 km s$^{-1}$ respectively for the working sunspot model by Maltby et al. (1986). Panels (A), (B) and (C) are as described in Fig.1.

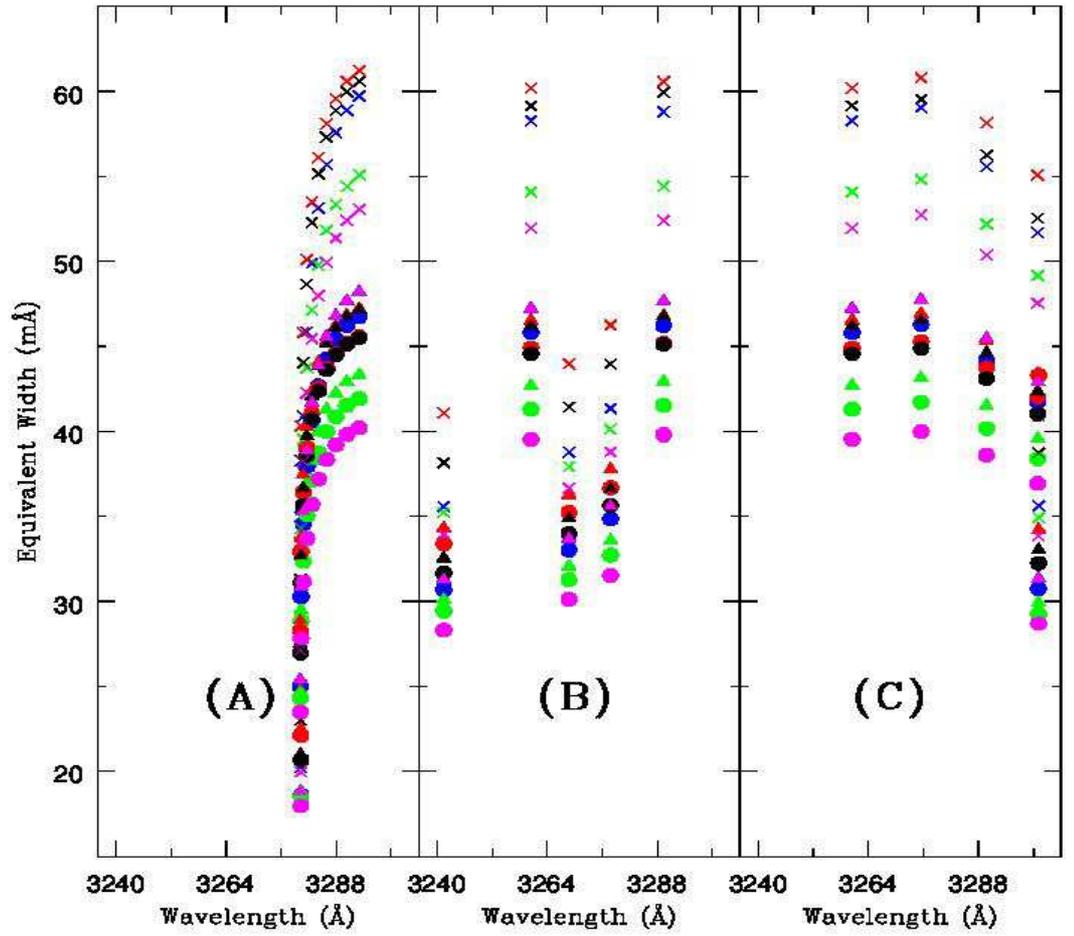

**Fig.4.** Predicted equivalent widths for isotopic lines of SH molecules in sunspot models plotted against wavelengths. Color code : Red, Green, Blue, Black and Magenta refer to sunspot models due to Avrett (1981), Boyer (1980), Zwaan (1974), Stellmacher & Weihr (1975) and Hennoux (1969) models respectively. Filled circles, filled triangles and crosses refer to micro-turbulence values as 0.85, 1.00 and 2.1 km s$^{-1}$ respectively. Panels (A), (B) and (C) are as described in Fig.1.